\documentclass[10pt,english,,cspaper,compsoc]{IEEEtran}
\usepackage[T1]{fontenc}
\usepackage[latin9]{inputenc}
\usepackage{amsthm}
\usepackage{amsmath}
\usepackage{amssymb}
\usepackage{graphicx}
\usepackage{esint}

\makeatletter
\theoremstyle{plain}
\newtheorem{thm}{\protect\theoremname}
\theoremstyle{definition}
\newtheorem{example}[thm]{\protect\examplename}
\theoremstyle{definition}
\newtheorem{defn}[thm]{\protect\definitionname}
\theoremstyle{remark}
\newtheorem{rem}[thm]{\protect\remarkname}
\theoremstyle{remark}
\newtheorem{claim}[thm]{\protect\claimname}
\theoremstyle{plain}
\newtheorem{lem}[thm]{\protect\lemmaname}

\ifCLASSOPTIONcompsoc
\else
\fi
\ifCLASSINFOpdf
\else
\fi
\makeatother

\usepackage{babel}\providecommand{\claimname}{Claim}
\providecommand{\definitionname}{Definition}
\providecommand{\examplename}{Example}
\providecommand{\lemmaname}{Lemma}
\providecommand{\remarkname}{Remark}
\providecommand{\theoremname}{Theorem}

\makeatother

\usepackage{babel}
\providecommand{\claimname}{Claim}
\providecommand{\definitionname}{Definition}
\providecommand{\examplename}{Example}
\providecommand{\lemmaname}{Lemma}
\providecommand{\remarkname}{Remark}
\providecommand{\theoremname}{Theorem}

\begin{document}

\title{On a Relation Between the Integral Image Algorithm and Calculus}

\author{Amir Shachar }
\maketitle
\begin{abstract}
The Integral Image algorithm is often applied in tasks that require
efficient integration over images, such as object detection. In this
paper we discuss theoretical aspects of the algorithm's continuous
version. We suggest to define the coefficients at the formulation
of the algorithm by applying a novel kind of discrete derivative.
Based on that operator we build a novel integration method over curves
in the plane, and apply it in a theorem that extends the algorithm
to general continuous domains.
\end{abstract}

\section{Previous Work}

\IEEEPARstart{E}{ver} since the early 1980's, computer scientists
have been using a formula named \textquotedbl{}Summed Area Table\textquotedbl{},
also known as \textquotedbl{}Integral Image\textquotedbl{}. The formula
allows a a rapid evaluation of the sum of rectangles in a given table
(or in an image), given that the 'image of sums', or the Integral
Image, is pre-evaluated. The formula was first introduced in 1984
by Crow (in \cite{crow}), and was reintroduced to the computer vision
community in 2001 by Viola and Jones (in \cite{viola_jones}).

The formula is detailed below. Given a function $i$ over a discrete
domain $\underset{{\scriptscriptstyle j=1}}{\overset{{\scriptscriptstyle 2}}{\prod}}\left[m_{j},M_{j}\right]\subset\mathbb{\mathbb{Z}}^{2}$,
define a new function $I$: 
\[
I\left(x,y\right)\equiv\overset{}{\underset{{\scriptscriptstyle x'\leq x\bigwedge y'\leq y}}{{\displaystyle \sum}}}i\left(x',y'\right),
\]
 and now the sum of all the values that the function $i$ accepts
on the grid $\left[a,b\right]\times\left[c,d\right],$ where $m_{1}\leq a,b\leq M_{1}$
and $m_{2}\leq c,d\leq M_{2}$, equals: 
\[
\overset{{\scriptscriptstyle b}}{\underset{{\scriptscriptstyle x'=a}}{{\displaystyle \sum}}}\overset{{\scriptscriptstyle d}}{\underset{{\scriptscriptstyle y'=c}}{{\displaystyle \sum}}}i\left(x',y'\right)=I\left(b,d\right)+I\left(a,c\right)-I\left(a,d\right)-I\left(b,c\right).
\]
This formula proved useful in the past few years, driving algorithms
such as Integral Video (see \cite{integral_video}), and a rotated
version of the Integral Image (see \cite{rotated_integral_image}).
Popular applications of the Integral Image formula, to name a few,
are efficient face detection (as performed in \cite{viola_jones}),
pedestrian detection (see \cite{pedestrians}), and Integral Histogram
(see \cite{integral-histogram}). Previously known algorithms had
also been enhanced via the Integral Image formula, such as the SIFT
algorithm (see \cite{integral_image_SIFT}). There have also been
works that suggest enhancements to the Integral Image formula itself,
such as Hensley et al.'s, see \cite{fast_summed_area}.

\begin{figure}
\enskip{}\includegraphics[scale=0.55]{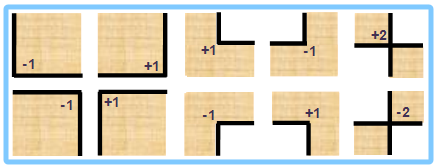}

\caption{\label{wang_corners}The corners that Wang et. al defined in their
paper. The number is the parameter $\alpha_{D}$, i.e., the coefficient
of the antiderivative in formula \ref{wang_theorem} for the case
$n=2$, at the specific corner. The location of the domain $D$ with
respect to the corner, is highlighted in brown.}
\end{figure}

In their work on applications of the Integral Image formula, Wang,
Doretto et al. (in \cite{wang} and \cite{doretto}) also suggested
a rigorous formulation to a natural extension of the formula as follows.
Let $D\subset\mathbb{R}^{n}$ be a domain that consists of a finite
unification of axis aligned rectangles - those whose edges are axis
aligned, and let $f:\mathbb{R}^{n}\rightarrow\mathbb{R}$ be an integrable
function. Let $F$ be an antiderivative of $f$, i.e., $F\left(\vec{x}\right)\equiv\underset{{\scriptscriptstyle B\left(\vec{x}\right)}}{\int}f\overrightarrow{dx}$,
where $B\left(\vec{x}\right)\subset\mathbb{R}^{n}$ is an axis aligned
box, that is determined according to the point $\vec{x}$ and the
origin. Then: 
\begin{equation}
\underset{{\scriptscriptstyle D}}{\int}f\overrightarrow{dx}=\underset{{\scriptscriptstyle \vec{x}\in\nabla\cdot D}}{\sum}\alpha_{D}\left(\vec{x}\right)\cdot F\left(\vec{x}\right),\label{wang_theorem}
\end{equation}
 where $\nabla\cdot D$ is the set of corners of the given domain
$D$, and $\alpha_{D}:\mathbb{R}^{n}\rightarrow\mathbb{Z}$ is a map
that depends on $n$. For $n=2$ it is such that $\alpha_{D}\left(\vec{x}\right)\in\left\{ 0,\pm1,\pm2\right\} $
according to which of the 10 types of corners, depicted in figure
\ref{wang_corners}, $\vec{x}$ belongs to. Note that formula \ref{wang_theorem}
extends the Integral Image formula in the sense that it is stated
for continuous domains ($\mathbb{R}$ rather than $\mathbb{Z}$),
and for more general types of domains (a finite unification of rectangles,
rather than plain rectangles).

\section{This Research's Goals}

Two theoretical questions rise from observing formula (\ref{wang_theorem}).
The first is, how can the coefficients $\alpha_{D}$ be defined given
a parametrization of the domain's edge, $\partial D$. The second
question is, how can this theorem be further extended to more general
types of domains (rather than finite unification of axis aligned boxes).
\textbf{Section 3} introduces a novel, semi-discrete pointwise operator,
namely a function's \textit{detachment}, which is in turn used for
a definition of the parameter $\alpha_{D}$ from formula \ref{wang_theorem},
given a parametrization of the domain's edge. The definition is coherent
in the sense that it is independent of the curve's different parametrizations.
\textbf{Section 4} suggests a natural extension of formula \ref{wang_theorem}
in the plane, to more general domains than finite unifications of
rectangles.

\section{Defining the Coefficients $\alpha_{D}$ By Classifying Corners}

\subsection{Corners Classifications Via The Curve's Derivatives}

Since the parameter $\alpha_{D}$ at formula (\ref{wang_theorem})
is uniquely defined according to the corner's type, then it is required
to introduce a tool for \textbf{corners classification} along a given
curve. Namely, given a curve $\gamma\left(t\right)=\left(x\left(t\right),y\left(t\right)\right)$,
we would like to point out a parameter that enables a proper distinction
between different types of corners along this curve and other curves,
as depicted in figure \ref{wang_corners}. Intuitively, given a corner
point $\gamma\left(t_{0}\right)$ along the curve, we would expect
the curve's one-sided derivatives vector at the corner point, that
is, $\left(x_{+}^{,},x_{-}^{,},y_{+}^{,},y_{-}^{,}\right)^{T}|_{{\scriptscriptstyle t=t_{0}}}$,
to gain a constant value for corners of a certain type and in this
sense be able to distinguish between different types of corners. The
next example clarifies in what sense this intuition is incorrect,
and the consequence will be that the curve's one-sided derivatives
vector is an incoherent tool in the task of corners classification. 
\begin{example}
\begin{figure}[h]
\enskip{}\enskip{}\enskip{}\enskip{}\enskip{}\enskip{}\enskip{}\enskip{}\enskip{}\enskip{}\enskip{}\enskip{}\enskip{}\enskip{}\includegraphics[scale=0.5]{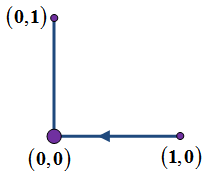}

\caption{\label{fig:derivative_not_coherent_illustration}An illustration to
the curve $C$, whose different parametrizations are discussed in
example \ref{derivative_not_coherent_example}.}
\end{figure}

\label{derivative_not_coherent_example}Let us analyze different parametrizations
of the same curve (see figure \ref{fig:derivative_not_coherent_illustration})
with a corner at $\left(0,0\right)$. Let us evaluate the curve's
one-sided derivatives at the corner point, for different parametrizations
differed by the value of $k$, 
\begin{eqnarray*}
 & C:\ \gamma_{k}\left(t\right),\;0\leq t\leq2\\
 & \gamma_{k}\left(t\right)=\begin{cases}
\left(\left(1-t\right)^{k},0\right), & 0\leq t\leq1\\
\left(0,\left(t-1\right)^{k}\right), & 1\leq t\leq2,
\end{cases}
\end{eqnarray*}
 where $k\in\mathbb{R}^{+}$. Note that the corner is accepted at
$t=1$.

For $k=1$, $\gamma_{1}$ forms an arc-length parametrization of the
curve. In that case, the curve's one-sided derivatives at the corner
point are $x_{+}^{,}\left(1\right)=0,\; x_{-}^{,}\left(1\right)=-1$
and: $y_{+}^{,}\left(1\right)=+1,\; y_{-}^{,}\left(1\right)=0$.

For $k\in\left(0,1\right)$, some of the one-sided derivatives of
$\gamma_{k}$ do not exist at the corner point (that is, $x_{-}^{,}\left(1\right),\; y_{+}^{,}\left(1\right)$
are undefined).

For $k>1$, the one-sided derivatives of $\gamma_{k}$ are all zeroed
at the corner point. 
\end{example}
The consequence from example 1 is that the vector: 
\[
\left(x_{+}^{,},x_{-}^{,},y_{+}^{,},y_{-}^{,}\right)^{T}|_{{\scriptscriptstyle t=t_{0}}}
\]
 is not a coherent tool in the task of corners classification, since
this vector is dependent of the curve's parametrization. A possible
approach to resolve that lack of consistency is to assume an arc-length
parametrization whenever it is required to classify a corner. However,
this approach ignores uncountably many other parametrizations of the
curve that are left unhandled. Hence, a different tool is required. 

Notice that in fact, the derivative inquires superfluous information
for this task, since we can settle for less information and inquire
the \textbf{sign} of the one-sided derivatives of the functions that
form the curve's parametrization. However, a similar analysis shows
that the vector of the one-sided derivatives' signs is not a coherent
tool either. 

In sub-section \ref{sub_section_detachment} we introduce a simple
tool, whose definition results from the following question: why calculate
the curve's rate of change to begin with, if all we are interested
in - is its trend of change?

\subsection{\label{sub_section_detachment}Definition of a Function's Detachment}

\begin{figure}[h]
\enskip{}\includegraphics[scale=0.19]{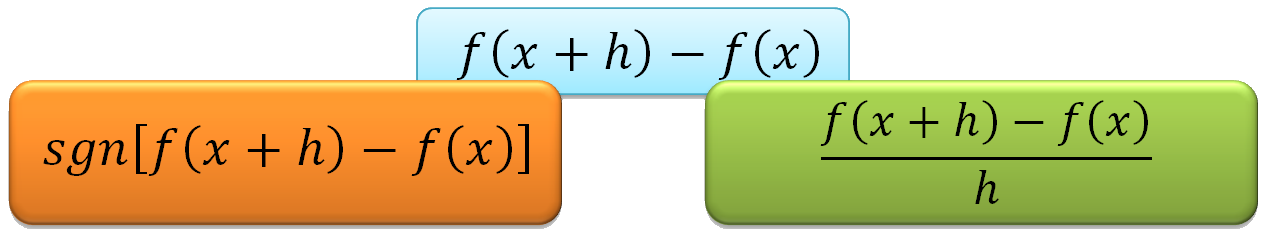}

\caption{\label{fig:derivative_vs_detachment}Let us observe the change in
the value of the function, $f\left(x+h\right)-f\left(x\right)$. It
is counter-productive to apply the limit process directly to that
term, since for any continuous function it holds that $\protect\underset{{\scriptscriptstyle h\rightarrow0^{\pm}}}{\lim}\left[f\left(x+h\right)-f\left(x\right)\right]=0$.
The derivative, however, manages to supply information regarding the
function's local rate of change by inquiring a different type of information:
it compares $dy$ and $dx$, via a fraction. The detachment is satisfied
with less information, and $dy$ is quantized, via the $sgn\left(\cdot\right)$
function. A function's detachment reveals a superficial information
regarding the function's instantaneous trend of change.}
\end{figure}

In order to illustrate the incoherency that rises from example \ref{derivative_not_coherent_example}
in a clearer manner, let us apply a relaxation to this problem, and
transform it from a problem on curves, to a problem on single variable
monomials.
\begin{example}
\label{function_derivative_incoherent}Let us consider the following
family of monomials: 
\begin{eqnarray*}
 & f_{k}:\;\mathbb{R}^{+}\rightarrow\mathbb{R}\\
 & f_{k}\left(x\right)=x^{k},
\end{eqnarray*}
 where $k\in\mathbb{R}^{+}$ is a positive real number. The right
derivative at zero equals: 
\[
\left(f_{k}\right)_{+}^{,}\left(0\right)=\begin{cases}
\text{undefined}, & k\in\left(0,1\right)\\
+1, & k=1\\
0, & k>1.
\end{cases}
\]
 Thus, when applied at the point $x=0$, the function's right derivative
depends on the value of $k$, and vanishes (either undefined or zeroed)
in uncountably many cases.

Let us introduce a pointwise operator which is robust in the sense
that it is independent of the parameter $k$. This operator supplies
a scant (yet sufficient for our requirements) amount of information
regarding the function's local monotony behavior.\end{example}
\begin{defn}
\emph{\label{def:Detachment}Detachment of a function}. Let us define
the one-sided detachments of a function at a point as: 
\begin{eqnarray*}
 & f_{\pm}^{;}:\ \mathbb{R}\rightarrow\left\{ 0,\pm1\right\} \\
 & f_{\pm}^{;}\left(x\right)\equiv\underset{{\scriptscriptstyle h\rightarrow0^{\pm}}}{\lim}sgn\left[f\left(x+h\right)-f\left(x\right)\right],
\end{eqnarray*}
 if the one-sided limits exist. The definition is illustrated in figures
\ref{fig:derivative_vs_detachment} and \ref{fig:Classification-of-corners}. 
\end{defn}

In terms of example \ref{function_derivative_incoherent}, it is verifiable
that applying one-sided right detachment to the function $f_{k}$
at zero equals: 
\[
\left(f_{k}\right)_{+}^{;}\left(0\right)=+1,
\]
 independently of the parameter $k$. Thus, the detachment is capable
of supplying an information regarding the local monotony behavior
of the function near a point also in cases where the derivative vanishes.
\begin{rem}
While the detachment may seem as a withered derivative at first glance,
this is not always the case, as shown at part 2 in \cite{applying_semi_discrete}.
Besides the theoretical advantage that we discuss here (as we saw,
the detachment can be thought of as an extension to the sign of the
derivative), the detachment may also be slightly more lightweight
than the derivative's sign in terms of computerized numerical approximation.
\end{rem}

\subsection{Corners Classifications Via The Curve's Derivatives}

\begin{figure}[h]
\includegraphics[scale=0.53]{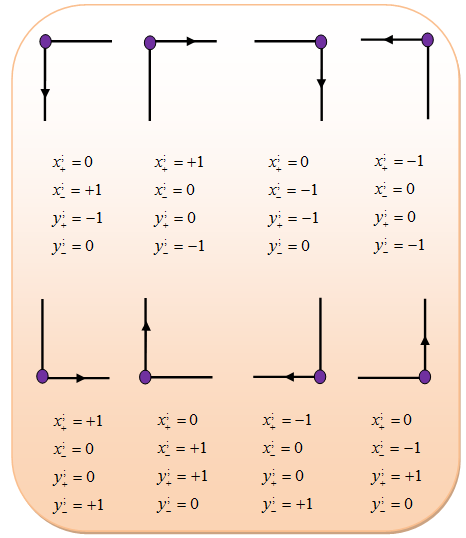}

\caption{\label{fig:Classification-of-corners}Classification of corners along
a curve according to the curve's one-sided detachments.}
\end{figure}

Going back to example \ref{derivative_not_coherent_example}, the
one-sided detachments vector: 
\[
\left(x_{+}^{;},x_{-}^{;},y_{+}^{;},y_{-}^{;}\right)^{T}|_{{\scriptscriptstyle t=1}}
\]
 equals $\left(0,+1,+1,0\right)^{T}$ regardless of the curve's parametrization,
and this vector distinguishes between different corners, as depicted
in figure \ref{fig:Classification-of-corners}.

The consequence is that the detachment is a coherent tool in the task
of corners classification, hence the coefficients $\alpha_{D}$ can
be defined via the one-sided detachments of the curve at the corner
point.

\section{Extending the Integral Image Algorithm to General Domains}

The formulation of the Integral Image algorithm in the plane, which
is formula \ref{wang_theorem} for the case where $n=2$, states a
relation between the double integral of a function over a finite unification
of axis aligned rectangles, and a linear combination of the given
function's antiderivative at the domain's corners. Since that formula
also extends part of the Fundamental Theorem of Calculus to $\mathbb{R}^{2}$,
we will abbreviate it FTC from now on.

A natural question that rises from this theorem is, how can it be
extended to more general types of domains, not merely those that are
formed by a finite unification of rectangular domains. Note that Pham
et al. (see \cite{extended_wang_2010}) suggested to extend the theroem
to polygonial domains via dynamic programming. We will seek to extend
it to any general domain in the plane, and not necessarily polygonial.

This research suggests that to extend the FTC, it is required to introduce
a novel integration method over curves in the plane. A by-product
of this integration method is a division of the domain bounded by
the curve - into a finite unification of axis aligned rectangles (for
whom the FTC is applied) and 'all the rest', for whom the double integral
of the original function is evaluated separately. Let us build this
integration method, step by step. Note that the following definitions
and claims are brought here in a concise manner. For a deeper discussion
on the intuition behind the definitions and a full proof of the claims,
see part 3 in \cite{applying_semi_discrete}.

\subsection{Definition of a Curve's Detachment}

In this subsection we define a term that enables to extend the coefficients
$\alpha_{D}$ from theorem \ref{wang_theorem} to any point on the
curve and not merely to corners - namely, a curve's detachment.
\begin{defn}
\emph{\label{curve-detachment-definition}Detachments vector of a
curve}. Let $C$ be a curve and let $\gamma\left(t\right)=\left(x\left(t\right),y\left(t\right)\right)$
be any parametrization of $C$, where $t\in\left(\alpha,\beta\right)$.
Let $z=\gamma\left(t_{0}\right)\in C$ be a point on the curve. If
the curve's one-sided detachments at $t_{0}$ (see definition \ref{def:Detachment}),
$x_{+}^{;}\left(t_{0}\right),\; x_{-}^{;}\left(t_{0}\right),\; y_{+}^{;}\left(t_{0}\right),\; y_{-}^{;}\left(t_{0}\right),$
all exist, then the curve is said to be detachable at $t_{0}$, and
the curve's detachments vector there is the vector: 
\[
\left(x_{+}^{;},x_{-}^{;},y_{+}^{;},y_{-}^{;}\right)^{T}|_{{\scriptscriptstyle t_{0}}}\in\left\{ 0,\pm1\right\} ^{4}.
\]

\end{defn}
For the simplicity of the discussion, we assume that the curves we
will discuss are detachable, continuous and simple. 
\begin{defn}
\textit{\label{Detachment-of-a-curve}Detachment of a curve}. Let
$C$ be a detachable curve in the plane and let $\gamma\left(t\right)=\left(x\left(t\right),y\left(t\right)\right)$
be any parametrization of $C$, where $t\in\left(\alpha,\beta\right)$.
We denote the detachment of the curve at a point $z=\gamma\left(t_{0}\right)\in C$
on the curve by $C^{;}\left(t_{0}\right)$, and define it as a function
of the curve's detachments vector as follows: 
\begin{eqnarray*}
 & C^{;}:\; C\rightarrow\left\{ 0,\pm1\right\} \\
 & C^{;}\left(z\right)\equiv y_{-}^{;}sgn\left(y_{-}^{;}-x_{-}^{;}\right)-y_{+}^{;}sgn\left(y_{+}^{;}-x_{+}^{;}\right).
\end{eqnarray*}

\end{defn}
\begin{figure}
\includegraphics[scale=0.62]{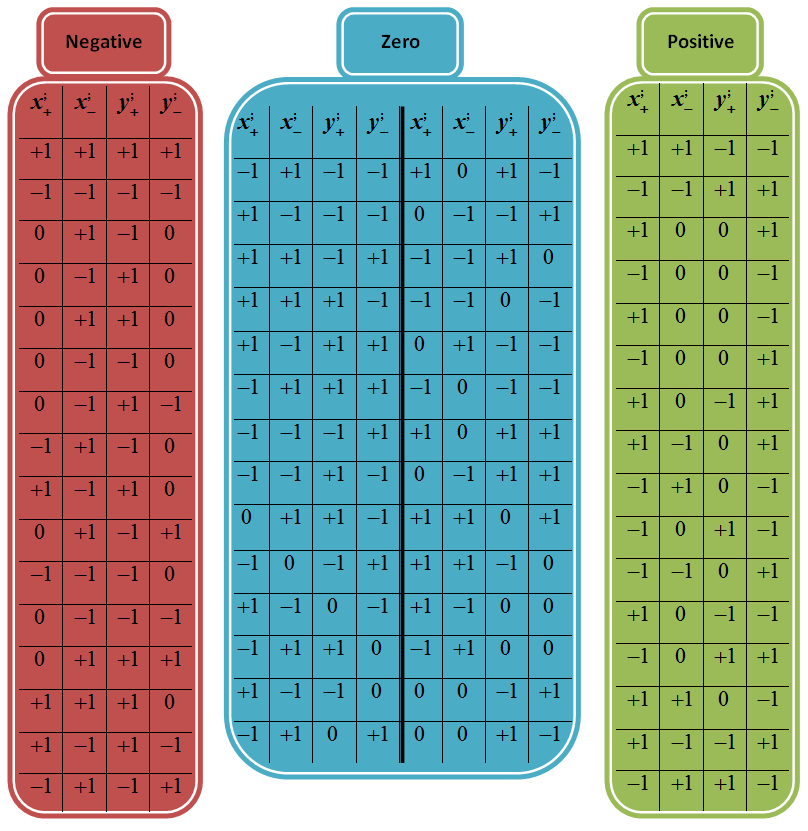}

\caption{\label{fig:A-curve's-detachment}The values of a curve's detachment
(definition \ref{Detachment-of-a-curve}) as a function of the curve's
detachments vector. The table should read as follows: Positive and
negative stand for $+1$ and $-1$ respectively, and if the curve's
detachments vector at a point is $\left(+1,+1,+1,+1\right)^{T}$,
$\left(-1,+1,-1,-1\right)^{T}$ or $\left(+1,+1,-1,-1\right)^{T}$,
then the curve's detachment is $-1$,$0$, or $+1$, respectively.}
\end{figure}

Note that the curve's detachment agrees with the coefficients $\alpha_{D}$
from the FTC along a curve's corners, and it extends it in the sense
that it is defined also at non-corner points. Further, a curve's detachment
is robust in the sense that it is independent of the curve's parametrization.
The general geometric interpretation of the curve's detachment is
introduced at part 3 in \cite{applying_semi_discrete}.

\subsection{Definition of Discrete Line Integral}

Equipped with the definition of a curve's detachment, let us now establish
the following integration method, whose aim is to extend the FTC to
more general domains than finite unifications of rectangles. Let us
first define the following terms. 
\begin{defn}
\emph{\label{def:monotonic-Curve.}Monotonic Curve}. A curve is said
to be monotonic if its detachments vector (see definition \ref{curve-detachment-definition})
is constant for each of its interior points. 
\end{defn}

\begin{defn}
\emph{\label{def:Positive-Domain-of}Positive Domain of a monotonic
curve}. Given a monotonic curve $\gamma$, let us define its positive
domain, $D^{+}\left(\gamma\right)$, as the domain bounded by the
curve and two axis-aligned lines, such that the domain is contained
in a left hand-side of the curve. 
\end{defn}
Definitions \ref{def:monotonic-Curve.} and \ref{def:Positive-Domain-of}
are illustrated in figure \ref{figure_of_lemma_for_extended_discrete}:
$ONO'$ is the positive domain of the monotonic curve $\gamma_{1}$.
\begin{defn}
\label{monotonic_division}Let $C$ be a detachable curve in $\mathbb{R}^{2}$.
A monotonic division of $C$ is an ordered set $\left\{ \left(\gamma_{i},\delta_{i}\right)\right\} _{1\leq i\leq n}$,
such that each $\gamma_{i}$ is a monotonic subcurve of $C$ whose
detachment is $\delta_{i}$, and $C=\underset{{\scriptscriptstyle 1\leq i\leq n}}{\bigcup}\gamma_{i}$.
\end{defn}

\begin{defn}
\emph{\label{definition:  Discrete-Line-Integral_monotonic}Discrete
Line Integral over a Monotonic Curve}. Let $f:\;\mathbb{R}^{2}\rightarrow\mathbb{R}$
be an integrable function, and let $F$ be its antiderivative. Let
$\gamma$ be a monotonic curve, contained in another curve, $\Gamma\supset\gamma$.
Then we define the discrete line integral of $F$ along the curve
$\gamma$ in the context of the curve $\Gamma$ as follows: 
\begin{eqnarray*}
\underset{{\scriptscriptstyle \gamma\subset\Gamma}}{\fint}F & \equiv & \underset{{\scriptscriptstyle D^{+}\left(\gamma\right)}}{\int\int}f\overrightarrow{dx}-\gamma^{;}F\left(B\right)\\
 &  & +\frac{1}{2}\left[\Gamma^{;}\left(A\right)F\left(A\right)+\Gamma^{;}\left(C\right)F\left(C\right)\right],
\end{eqnarray*}
 where $\gamma^{;}$ is the monotonic curve's detachment, $D^{+}\left(\gamma\right)$
is the given curve's positive domain, and $\Gamma^{;}\left(A\right),\;\Gamma^{;}\left(C\right)$
are the detachments of the curve $\Gamma$ at the points $A$ and
$C$ respectively along the edge of the domain.
\end{defn}

\begin{defn}
\emph{\label{def:Discrete-Line-Integral_detachable}}Let $C$ be a
detachable curve, and let $\left\{ \left(\gamma_{i},\delta_{i}\right)\right\} _{1\leq i\leq n}$
be a monotonic division of a subcurve $\gamma\subset C$. Let us consider
a function $f:\;\mathbb{R}^{2}\rightarrow\mathbb{R}$ that admits
an antiderivative $F$. Then the \textbf{discrete line integral of
$F$ over $\gamma$ in the context of the curve }$C$ is defined as
follows:
\[
\underset{{\scriptscriptstyle \gamma\subset C}}{\fint}F\equiv\underset{{\scriptstyle i}}{\sum}\underset{{\scriptscriptstyle \gamma_{i}\subset C}}{\fint}F,
\]
where each $\underset{{\scriptscriptstyle \gamma_{i}\subset C}}{\fint}F$
is calculated according to the definition of the discrete line integral
over monotonic curves, see definition \ref{definition:  Discrete-Line-Integral_monotonic}.
\end{defn}

\subsection{Algebraic Properties of Discrete Line Integral}

The definition of the discrete line integral becomes clearer as soon
as its algebraic properties are being proven. See \cite{applying_semi_discrete},
section 12, for a detailed discussion (and detailed proofs) of the
following properties.
\begin{claim}
\label{cla:discrete_integral_additivity}Let $\gamma$ be a monotonic
curve contained in another curve, $\Gamma\supset\gamma$, and let
$\gamma_{1}$ and $\gamma_{2}$ be its sub-curves, such that $\gamma=\gamma_{1}\bigcup\gamma_{2}$.
Then: 
\[
\underset{{\scriptscriptstyle \gamma\subset\Gamma}}{\fint}F=\underset{{\scriptscriptstyle \gamma_{1}\subset\Gamma}}{\fint}F+\underset{{\scriptscriptstyle \gamma_{2}\subset\Gamma}}{\fint}F.
\]
This claim also asserts that the discrete line integral over a detachable
curve (see definition \ref{def:Discrete-Line-Integral_detachable})
is indeed well defined, because it is independent of the monotonic
division of the curve. 
\end{claim}

\begin{claim}
\label{cla:discrete_line_integral_negativity}Let $\gamma$ be a monotonic
curve whose detachment is never zeroed, that is contained in another
curve, $\Gamma$. Then: 
\[
\underset{{\scriptscriptstyle \gamma\subset\Gamma}}{\fint}F=-\underset{{\scriptscriptstyle -\gamma\subset-\Gamma}}{\fint}F,
\]
 where $-\gamma,-\Gamma$ are the curves $\gamma,\Gamma$ (respectively)
with a flipped orientation. 
\end{claim}

\subsection{Applying the Discrete Line Integral to Extend the Integral Image
Algorithm}

A detailed proof of the following claims is available at section 13
in \cite{applying_semi_discrete}. Let us begin with the following
lemma. 
\begin{lem}
\label{lemma_extened_discrete}Let $\left\{ \left(\gamma_{i},\delta_{i}\right)\right\} _{1\leq i\leq n}$
be a monotonic division of a closed and detachable curve $\gamma$.
Let $f:\,\mathbb{R}^{2}\rightarrow\mathbb{R}$ be a function that
admits an antiderivative $F$. Let $M,N,O$ be the endpoints of the
curves $\gamma_{1},\gamma_{2}$ respectively (where $N=\gamma_{1}\bigcap\gamma_{2}$).
Let:
\[
\alpha\equiv\gamma_{1}\bigcup\gamma_{2}\bigcup\overrightarrow{MO},
\]
and:
\[
\beta\equiv\underset{{\scriptscriptstyle i=3}}{\overset{{\scriptscriptstyle n}}{\bigcup}}\gamma_{i}\bigcup\overrightarrow{OM.}
\]
Then:
\[
\underset{{\scriptscriptstyle \gamma}}{\fint}F=\underset{{\scriptscriptstyle \alpha}}{\fint}F+\underset{{\scriptscriptstyle \beta}}{\fint}F.
\]

\end{lem}
\begin{figure}
\includegraphics[scale=0.7]{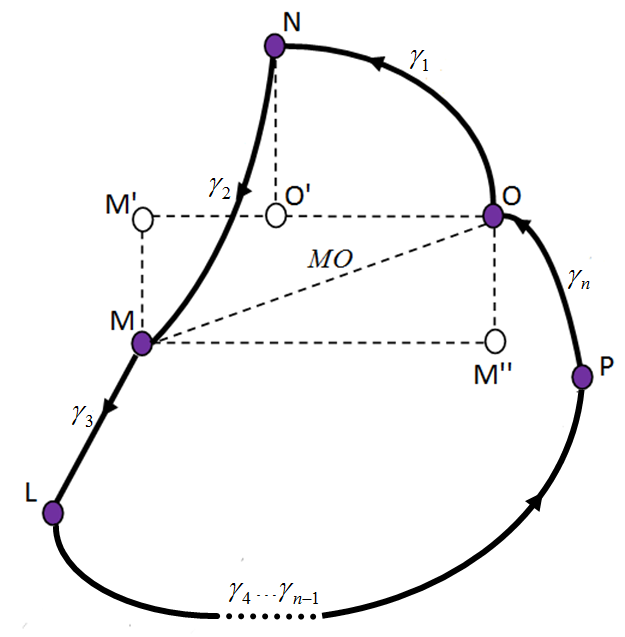}

\caption{\label{figure_of_lemma_for_extended_discrete}An illustration to lemma
\ref{lemma_extened_discrete}.}
\end{figure}

The lemma is illustrated in figure \ref{figure_of_lemma_for_extended_discrete}.
In short, the lemma's correctness follows by seperating to cases and
applying the definition of the discrete line integral.

\begin{figure*}
\enskip{}\includegraphics[scale=0.6]{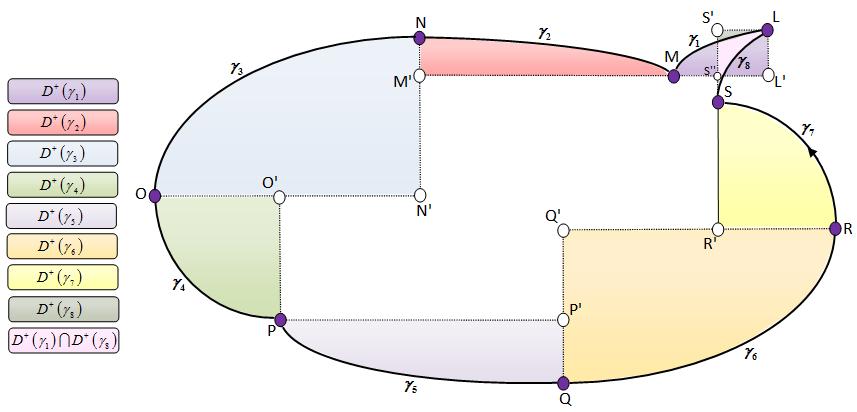}

\caption{\label{fig:extended_discrete_illustration}An illustration to example
\ref{exa:extended_discrete_example}. The edge of the domain is $\gamma\equiv\protect\underset{{\scriptscriptstyle 1\leq i\leq8}}{\bigcup}\gamma_{i}$,
and the positive domains of each sub-curve are colored according to
the legend on the left.}
\end{figure*}

\begin{thm}
\label{theorem_extended_discrete_green}Let $D\subseteq\mathbb{R}^{2}$
be a domain whose edge is detachable. Let $f:\mathbb{R}\rightarrow\mathbb{R}$
be a function that admits an antiderivative $F$. Then: 
\begin{equation}
\underset{{\scriptscriptstyle D}}{\int\int}f\overrightarrow{dx}=\underset{{\scriptscriptstyle \partial D}}{\fint}F.\label{eq:extended_discrete_green}
\end{equation}

\end{thm}
Formula \ref{eq:extended_discrete_green} can be shown to hold by
induction on the number of monotonic sub-curves that form the domain's
edge, $\partial D$. The induction's basis can be shown to hold by
seperating to cases of domains whose edge consists of merely 3 monotonic
subcurves. The induction's step can be done by applying both lemma
\ref{lemma_extened_discrete} and the induction's hypothesis. This
theorem is illustrated in \cite{wolf_extended_green}. The following
example demonstrates formula \ref{eq:extended_discrete_green}. 
\begin{example}
\label{exa:extended_discrete_example}Let $f:\mathbb{R}^{2}\rightarrow\mathbb{R}$
be an integrable function, and let $F:\mathbb{R}^{2}\rightarrow\mathbb{R}$
be its antiderivative. Let $\gamma$ be a curve as depicted in figure
\ref{fig:extended_discrete_illustration}. 
\end{example}
Let us calculate the discrete line integral over $\gamma$. First
we plug the detachments of the monotonic subcurves $\left\{ \gamma_{i}\right\} _{i=1}^{n}$
into the definition of the discrete line integral over monotonic curves
(according to definition \ref{definition:  Discrete-Line-Integral_monotonic},
for example $\underset{{\scriptscriptstyle \gamma_{1}}}{\fint}F=\underset{{\scriptscriptstyle D\left(\gamma_{1}\right)}}{\iint}f\overrightarrow{dx}-\frac{1}{2}F\left(L\right)+F\left(L'\right)$).
Then we add up the equations and deduct the twice-calculated double
integral over the rectangle $LS'S''L'$ due to the equation$\underset{{\scriptscriptstyle LS''S''L'}}{\int\int}f\overrightarrow{dx}=F\left(S\right)-F\left(S'\right)+F\left(S''\right)-F\left(L'\right)$
that follows from the FTC. The calculation results with $\underset{{\scriptscriptstyle \gamma}}{\fint}F\equiv\underset{{\scriptscriptstyle i=1}}{\overset{{\scriptscriptstyle n}}{\sum}}\underset{{\scriptscriptstyle \gamma_{i}}}{\fint}F=\underset{{\scriptscriptstyle D}}{\int\int}f\overrightarrow{dx}$,
as stated by formula \ref{eq:extended_discrete_green}. 
\begin{rem}
Note that formula \ref{eq:extended_discrete_green} extends formula
\ref{wang_theorem} in the sense that if $D$ is a finite unification
of axis aligned rectangles, then the discrete line integral over its
monotonic sub-curves consists of a linear combination of the antiderivative
alone, where the double integral over the positive domain is omitted,
because the positive domain of each such sub-curve is degenerated.
\end{rem}

\section{Summary}

Examining the continuous version of the Integral Image formula, we
have researched the following directions. First, we suggested that
a curve's detachments vector is an appropriate tool for corners classification,
and as such it enables to properly define the coefficients at the
formula. Second, we applied a novel discrete integration over curves
to generalize the formula - from rectangular to general domains. Those
tools may also have applicational advantages: a slight optimization
to the approximation of a derivative's sign due to the detachment,
and parallel integration over domains in the plane due to a discretization
of formula \ref{eq:extended_discrete_green}.

\end{document}